# Evaluation of efficiency index of friction energy dissipation devices using endurance time method

Amir Shirkhani *, Bahman Farahmand Azar **, Mohammad Charkhtab Basim ***

**Abstract:**

Various methods have been presented to improve the performance of buildings against earthquakes. Friction damper device is one of the energy dissipation devices that appropriately absorbs and dissipates the input energy and decreases displacements. In this paper, the possibility of using endurance time method to determine the efficiency index and optimum slip load for these dampers was investigated by comparing the results of endurance time and nonlinear time history analyses. The efficiency indexes acquired from the average of results for nonlinear time history and endurance time analyses were close to each other. In this research, by assuming identical optimum slip load for the dampers in all stories, the normalized damper strength was increased in a number of equal steps ranging from zero to one to determine the efficiency index of dampers in each step. Then, the optimum slip load of dampers in the steel frames was calculated according to the minimum efficiency index of dampers. As a result, employing the endurance time method instead of a high number of nonlinear time history analyses is also possible, and using the endurance time method diminishes 57% of computational endeavors. Lastly, a relation for acquiring the optimum slip load of the friction damper devices in steel structures was proposed in terms of the weight of the structures. After adding optimal FDDs to the structures and investigating the effectiveness of the dampers, it was concluded that by using endurance time excitation function with better energy consistency, the endurance time results could be improved.

## 1. Introduction

Unavoidable earthquakes that lead to a lot of financial and life losses necessitate finding a safe way against this natural event [1]. In the last decades, Passive control systems have successfully been employed to diminish the dynamic response of structures against earthquakes and violent winds. One of these systems is friction dampers since they offer the high potential of energy dissipation at a relatively low cost and are simple to install and maintain [2]. Mualla and Belev [2, 3] designed this Friction Damper Device (FDD). Nielsen and Mualla [4] determined a relatively precise bilinear approximation to define the behavior of a central damping system. Naeem and Kim [5] developed a friction damper with a restoring force by a combination of a torsional spring and an FDD. They showed that there is the least probability that the structure retrofitted with the proposed damper will reach the special limit states. Nabid et al. [6] developed an adaptive performance-based optimization approach for the optimal design of friction wall dampers in RC structures subjected to seismic excitations. Since the conventional Nonlinear Time History (NTH) requires much time to scale the earthquake records and perform a lot of analyses, the Endurance Time (ET) method was utilized. This method is successfully used in various studies [7-10]. The success of performing the ET method depends on the correct generation of the ET excitation functions (ETEFs). The ETEFs should be created in such a way they can be employed to anticipate the effects of real ground motion [11-13].

In this research, the optimum slip load of FDDs was calculated according to their Efficiency Index (EFI) in steel frames. The validation of the ET method is another objective of this study that was performed by comparing NTH and ET results. Finally, a relation for calculating the optimum slip load of the FDDs in steel frames was proposed in terms of the weight of the structure.

* PhD Candidate, Department of Structural Engineering, Faculty of Civil Engineering, University of Tabriz, Tabriz, Iran.
** Corresponding Author: Associate Professor, Department of Structural Engineering, Faculty of Civil Engineering, University of Tabriz, Tabriz, Iran. Email: b-farahmand@tabrizu.ac.ir
*** Assistant Professor, Faculty of Civil Engineering, Sahand University of Technology, Tabriz, Iran.

## 2. Friction Damper Device

The configuration of FDD is exhibited in Fig. 1. The FDD comprises a central plate, two side plates, and two circular friction pad discs located between the plates [14, 15]. The central plate of FDD is connected to the middle of the beam by a hinge so that it can freely displace to dissipate energy. Steel bracing bars are pin-connected at both of their ends to the base of the column and the FDD [2, 3].

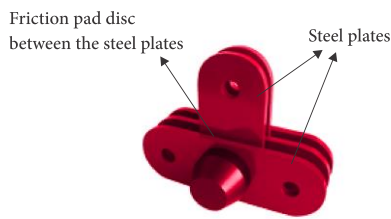

**Fig. 1:** Configuration of FDD [14]

To have a greater recognition of this damper, a one-story frame equipped with an FDD is shown in Fig. 2. The FDD includes the rigid plate $C_1CC_2$ of length $2r$ and AC of length $h_a$ and a frictional hinge C with the rotational frictional strength $M_f$ which equals to $h_a$ multiplied by slip load $F_h$ [4].

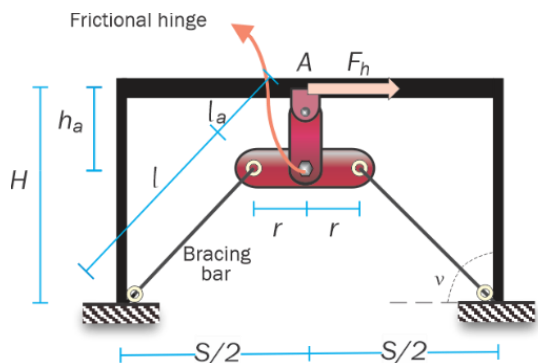

**Fig. 2:** Frame equipped with FDD

## 3. Endurance Time method

ET method is a dynamic pushover analysis procedure for anticipating structures' seismic performance in which the structures they are subjected to predefined intensifying dynamic excitations [16-19]. The predesigned excitations i.e. ETEFs are created so that their intensity is increased in conjunction with time. The performance of the structure in different Intensity Measures (IMs) is assessed by unit analysis in the ET method. Hence, some analyses are reduced significantly [20]. The ETEFs provide a proper evaluation of real responses of structures to utilize in Performance-based earthquake engineering (PBEE), provided, the major features of earthquake records assemble with various intensities to generate them [21]. One of the steps in the ET method is the selection of a set of ETEFs that corresponds with the desired site specifications. The template spectrum to produce ETEFs must match site-specific design spectra to most compatible with codified procedures. This can be attained by choosing ETEFs that have a similar template spectrum with either a code design spectrum or an average response spectrum of a suite of earthquake records corresponding to the specific site [20]. Three ETEFs that were employed in this study (ETA20f set) have been created in a manner so that their response spectrum is proportional to the average response spectrum of seven earthquake records from Table 1 [22]. The average response spectrum of seven scaled records, and "ETA20f set" excitation function and its response spectra at various times are illustrated in Figs. 3 and 4, respectively.

Seven recorded ground motions [23] for soil "type C" which have been used to produce "ETA20f set" excitations functions [13, 17], were selected to perform NTH analyses. As advised in the reference [8], three ETEFs in this set, which have only various optimization start points, can be utilized to decrease the random scatter influences in the results [9]. It is be noted that this study was basically performed by ETA20f series of ETEFs and this issue is stated as a limitation of this study with suggestions for further studies.

**Table 1:** Earthquake records set used in this research [23].

| Date | Earthquake name | Magnitude (Ms) | Station number | Component (deg) | PGA (g) | Abbreviation |
|---|---|---|---|---|---|---|
| 06/28/92 | Landers | 7.5 | 12 149 | 0 | 0.171 | LADSP000 |
| 10/17/89 | Loma Prieta | 7.1 | 58 065 | 0 | 0.512 | LPSTG000 |
| 10/17/89 | Loma Prieta | 7.1 | 47 006 | 67 | 0.357 | LPGIL067 |
| 10/17/89 | Loma Prieta | 7.1 | 58 135 | 360 | 0.450 | LPLOB000 |
| 10/17/89 | Loma Prieta | 7.1 | 1 652 | 270 | 0.244 | LPAND270 |
| 04/24/84 | Morgan Hill | 6.1 | 57 383 | 90 | 0.292 | MHG06090 |
| 01/17/94 | Northridge | 6.8 | 24 278 | 360 | 0.514 | NRORR360 |

13

## 4. Characteristics of steel frames

To compare the responses of steel moment frames equipped with and without FDDs, 2D regular generic frames with 3, 7, and 12 stories were considered here. The structures were designed on the basis of Iranian Standard No. 2800 [24] and Iranian National Building Code (INBC part 10) [25] which are close to AISC ASD design code [26] for steel structures. The soil selected was type 2 from Iranian Standard No. 2800 which was matched with site class C from ASCE/SEI 41-06 [27]. The specifications of the designed frames are presented in Tables 2.

**Table 2:** Specifications of steel frames.

| Frame | Number of Stories | Weight (kN) | Design base shear (kN) | First mode mass participation (%) |
|---|---|---|---|---|
| MF03S | 3 | 1458 | 182 | 81·12 |
| MF07S | 7 | 3491 | 312 | 78·25 |
| MF12S | 12 | 6065 | 414 | 74·93 |

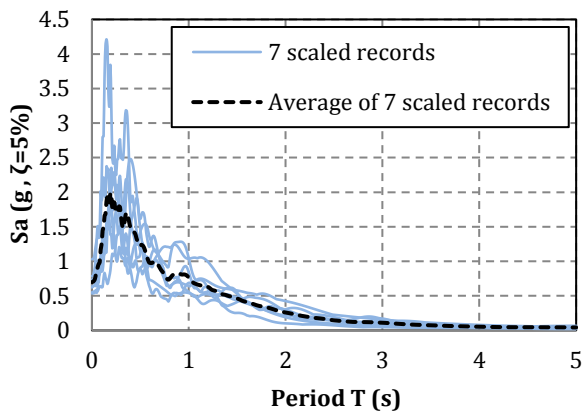

**Fig. 3:** Average response spectrum of seven scaled records

## 5. Modeling for analysis

To investigate the effect of FDD on seismic performance of steel moment frames, it was assumed that FDDs are situated in the middle spans of three-, seven- and 12-story steel frames. The diameter and yield stress of the bracing bars were determined to equal 3.5 cm and 355 MPa, respectively. The length of the central plate of FDDs was determined to equal 20 cm. Three-story steel frame equipped with FDDs is presented in Fig. 5. The FDD follows the Coulomb friction law; therefore, a rigid-plastic curve was employed for modeling the frictional hinge and the behavior of the FDD [28] as shown in Fig. 6. The period of free vibration of steel frames with and without FDDs is given in Table 3 by the prefix of DMF and MF, respectively. As it is obvious from Table 3, the period of free vibration diminished by supplementing FDDs to the structures.

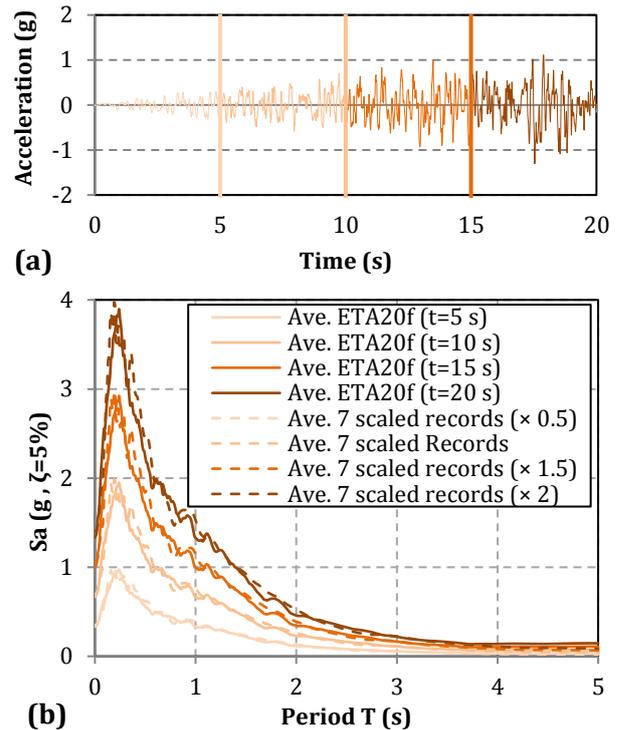

**Fig. 4:** (a) ETA20f01 excitation function; (b) acceleration response spectra at various times of excitation

## 6. Structural analysis

### 6.1. Scaling ground motions and calculating equivalent times of ETEFs

In this study, for scaling the records of the ground motions and calculating the equivalent time of ETEFs, the response spectrum of BSE-2 hazard level [27] was used. The earthquake records are scaled in a manner that their response spectra are situated above ASCE/SEI 41-06 spectrum for the period range 0.2Ti–1.5Ti, where Ti is the fundamental period of the structures. Since the structures are considered two-dimensional in this study, the records must be scaled individually rather than scaling them as pairs [8, 20]. The scale factors of earthquake records for NTH analysis of frames with and without FDD are presented in Table 4. The response spectrum for any window of the "ETA20f set" of ETEFs from to resembles that of the average response spectrum of the seven earthquake records with a scale factor that is proportionate with time. The scale factor equals unit value for  s in the present research [29]. To acquire the ET equivalent time for ETEFs, the spectrum utilized for creating them are scaled for each structure in the period range 0.2Ti–1.5Ti. The procedure for scaling is accurately alike to the one adopted for scaling earthquake records [29]. The ET equivalent times for the structures are listed in Table 5.

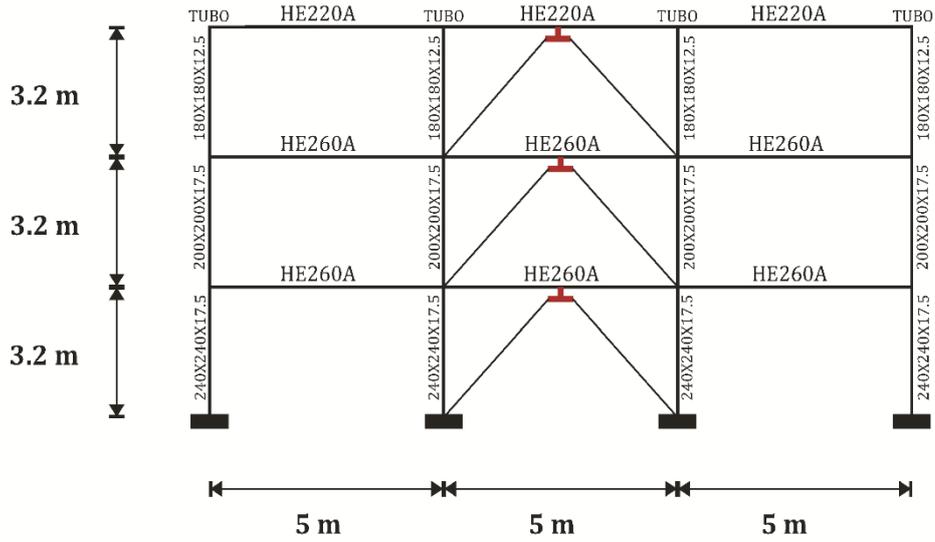

**Fig. 5:** Schematic of three-story frame equipped with FDDs

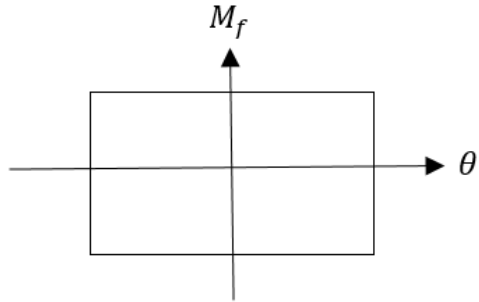

**Fig. 6:** Rigid-plastic curve for frictional hinge

**Table 3:** Period of free vibration of frames with and without FDDs.

| Frame | Period of free vibration (s) |
|---|---|
| MF03S | 0.94 |
| DMF03S | 0.48 |
| MF07S | 1.42 |
| DM07S | 0.95 |
| MF12S | 2.02 |
| DMF12S | 1.54 |

**Table 4:** Scale factors of earthquake records employed in NTH analysis for frames with and without FDD.

| Frame | LADSP000 | LPAND270 | MHG 06090 | LPGIL067 | LPLOB000 | LPSTG000 | NRORR360 |
|---|---|---|---|---|---|---|---|
| MF03S | 3.98 | 2.98 | 1.75 | 2.62 | 2.73 | 1.85 | 1.09 |
| DMF03S | 3.87 | 2.49 | 2.22 | 2.14 | 2.11 | 2.02 | 1.32 |
| MF07S | 4.20 | 3.13 | 2.05 | 2.83 | 3.74 | 1.66 | 1.13 |
| DMF07S | 3.98 | 2.99 | 1.75 | 2.62 | 2.74 | 1.84 | 1.09 |
| MF12S | 4.78 | 3.65 | 2.78 | 3.47 | 5.46 | 1.71 | 1.31 |
| DMF12S | 4.33 | 3.27 | 2.20 | 2.94 | 4.08 | 1.65 | 1.15 |

**Table 5:** ET Equivalent times for the frames with and without FDDs.

| Frame | Equivalent time (s) |
|---|---|
| MF03S | 11.1 |
| DMF03S | 10.9 |
| MF07S | 12.3 |
| DMF07S | 11.1 |
| MF12S | 14.2 |
| DMF12S | 12.7 |

### 6.2. Determination of optimum slip load of FDDs

Determination of optimum slip load is the major key in the optimum design of FDD. The Efficiency Index (EFI) introduced by Mualla and Belev [2] for FDD is defined as follows:

$$EFI = \sqrt{R_d^2 + R_f^2 + R_e^2} \quad (1)$$

where $R_d$, $R_f$, and $R_e$ are the displacement diminution factor, the base shear diminution factor and the remained energy factor, respectively, which have been defined in reference [2, 30]. The remained energy factor can also be calculated as follows:

$$R_e = (100 - \text{Dissipated Energy})/100 \quad (2)$$

It is noteworthy that the best efficiency of FDD is when it is effective in the diminution of displacement, base shear, and remained energy. To this end, the normalized FDD strength is defined as follows:

$$\eta_M = M_f / M_u \quad (3)$$

where $M_f$ is the rotational frictional strength of FDD and $M_u$ is torque demand imposed on FDD by ground shaking whereas the damper is locked. In this research, by assuming identical optimum slip load for the FDDs in all stories, the normalized FDD strength was increased in 20 steps ranging from zero to one to determine the EFI of FDDs in each step and then, the optimum slip load of FDDs in the steel frames was calculated according to the minimum EFI of FDDs. The energy dissipated by FDDs and the EFI of FDDs for seven-story frame based on NTH and ET results are shown in Figs. 7-10. The performance of FDDs in the steel frames is illustrated in Fig. 11.

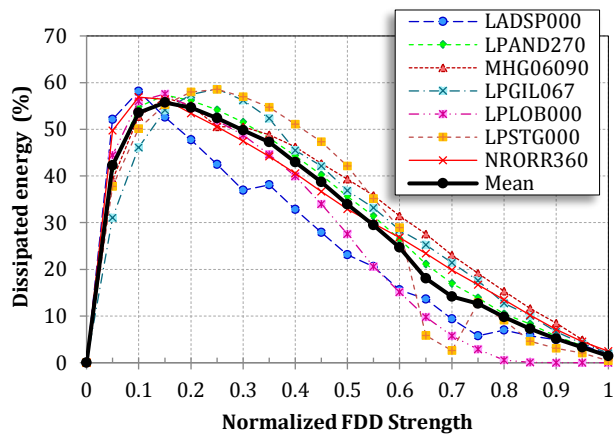

**Fig. 7:** Energy dissipated by FDDs in seven-story frame based on NTH results

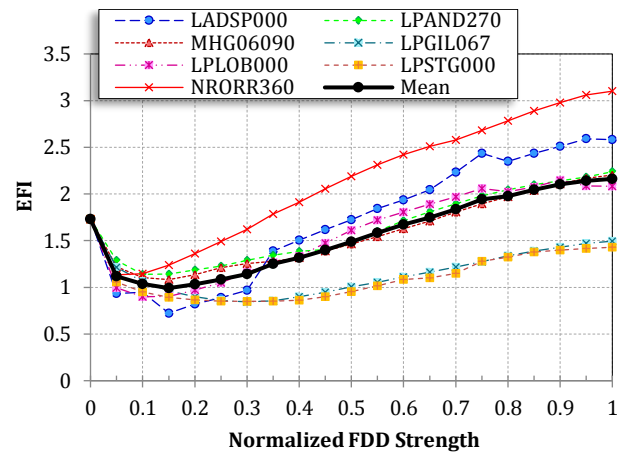

**Fig. 9:** EFI of FDDs in seven-story frame based on NTH results

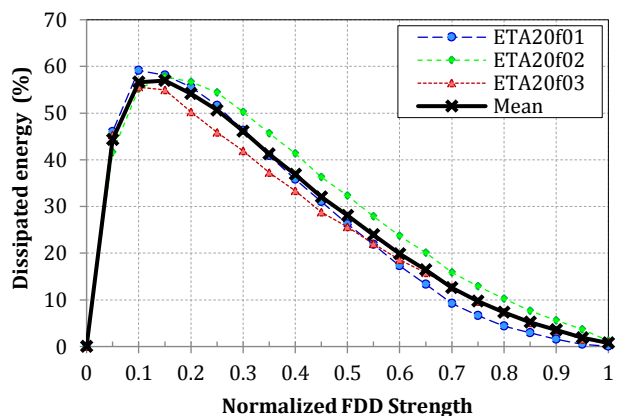

**Fig. 8:** Energy dissipated by FDDs in seven-story frame based on ET results

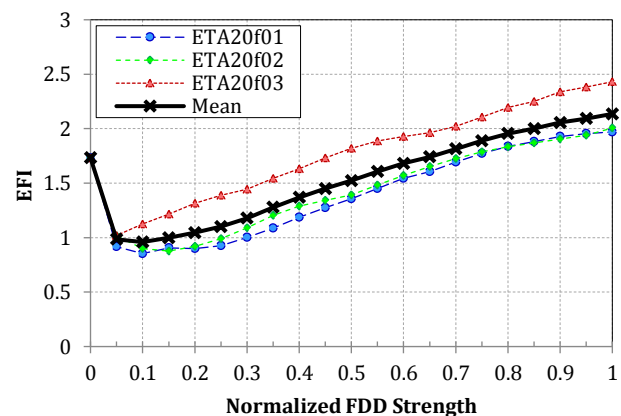

**Fig. 10:** EFI of FDDs in seven-story frame based on ET results

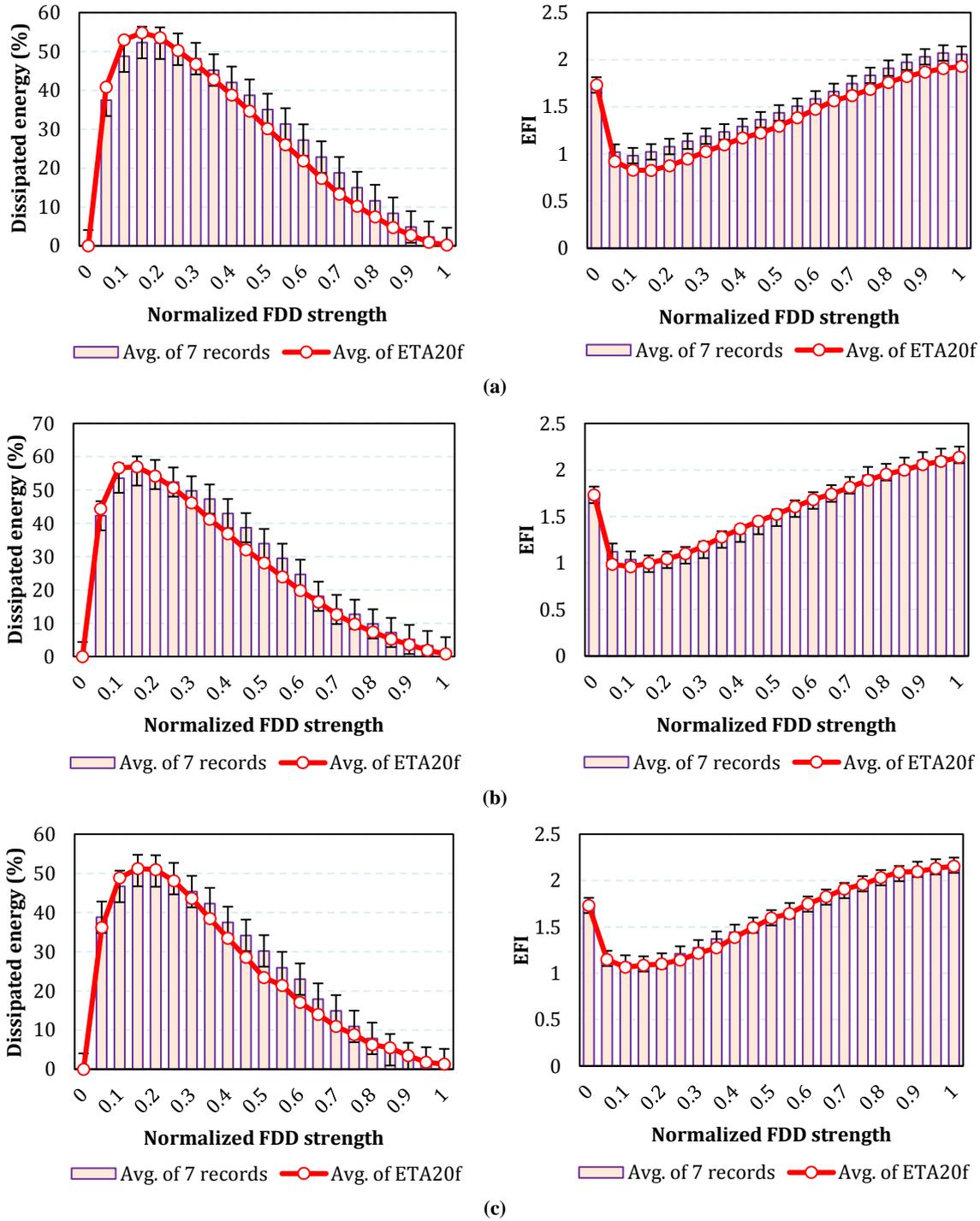

**Fig. 11:** Performance of FDDs in the frames: (a) three-story frame; (b) seven-story frame; (c) 12-story frame

According to Fig. 11, the minimum EFI of FDDs for the mean values of NTH and ET results are obtained in $\eta_{M,opt} = 0.1$ and $0.15$ for both analysis procedures in three-, seven- and 12-story frames. Table 6 present these values. The EFIs corresponding to $\eta_{M,opt} = 0.1$ and $0.15$ are close to each other for both NTH and ET analyses.

**Table 6:** Minimum EFI of FDDs for the mean values of NTH and ET results.

| Frame | Analysis | $\eta_M$ | EFI | Difference in EFI (%) |
|---|---|---|---|---|
| Three-story | NTH | 0.1 | 0.983 | 3.9 |
|  |  | 0.15 | 1.023 |  |
|  | ET | 0.1 | 0.828 | 0.1 |
|  |  | 0.15 | 0.827 |  |

17

| | | | | | |
|---|---|---|---|---|---|
| Seven-story | NTH | 0.1 | 1.035 | 4.2 | |
| | | 0.15 | 0.992 | | |
| | ET | 0.1 | 0.959 | 4.0 | |
| | | 0.15 | 0.999 | | |
| 12-story | NTH | 0.1 | 1.111 | 1.0 | |
| | | 0.15 | 1.100 | | |
| | ET | 0.1 | 1.065 | 2.0 | |
| | | 0.15 | 1.087 | | |

The percentages of energy dissipated by FDDs corresponding to $\eta_{M,opt} = 0.1$ and $0.15$ are also close to each other for both analysis procedures. Based on NTH and ET results and considering $\eta_{M,opt} = 0.1$, the following relations are proposed to determine the optimum slip load $F_{h,opt}$ of FDDs in the steel frames, as shown in Fig. 12.

$$F_{h,opt} = A_w.W \qquad (4)$$

where $W$ is the weight of the structure and coefficient $A_w$ is defined as follows:

$$\begin{cases} \text{for NTH analysis:} \\ A_w = (2\times 10^{-9})W^2 - (3\times 10^{-5})W + 0.1142; \\ \text{for ET method} \\ A_w = (2\times 10^{-9})W^2 - (3\times 10^{-5})W + 0.1079 \end{cases} \qquad (5)$$

According to the above equations, it can be observed that $A_w$ and consequently $F_{h,opt}$ are close to each other for both analysis procedures. Therefore, it can be deduced that the results of the ET method are valid.

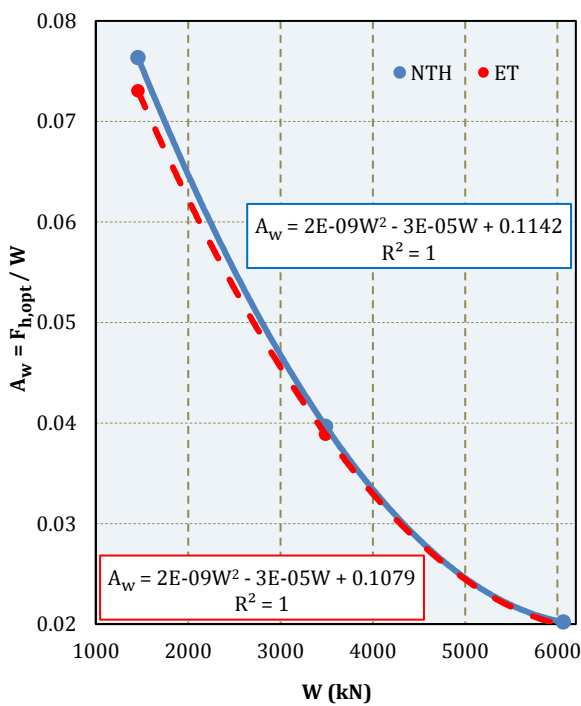

**Fig. 12:** Relation between $A_w$ and $W$

Figs. 13-15 shows the effectiveness of optimal FDDs in multi-story structures based on ET and NTH analyses. Fig. 15 indicates that by using ETEFs with better energy consistency, the ET results could be improved.

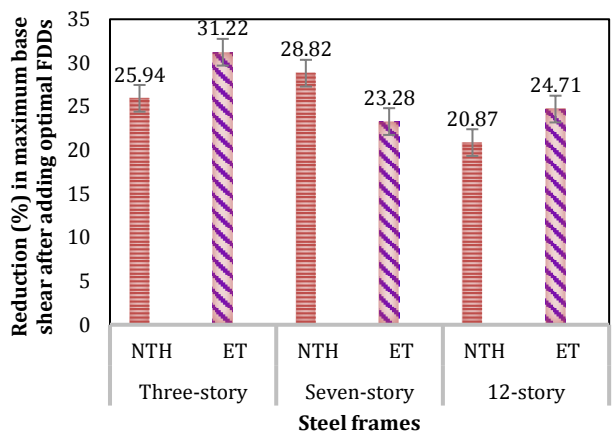

**Fig. 13:** Reduction (%) in maximum base shear after adding optimal FDDs based on ET and NTH results

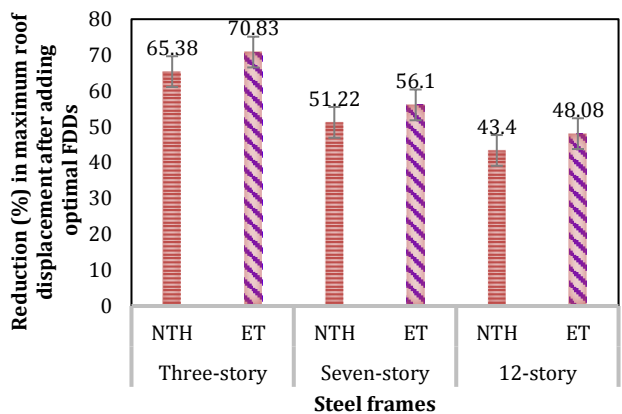

**Fig. 14:** Reduction (%) in maximum roof displacement after adding optimal FDDs based on ET and NTH results

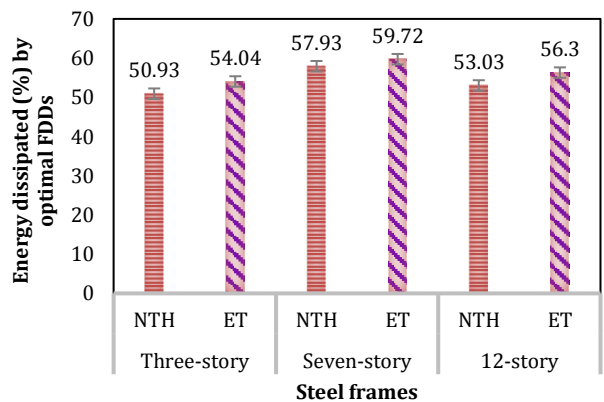

**Fig. 15:** Energy dissipated (%) by optimal FDDs based on ET and NTH results

## 7. Conclusions

This paper presents a practical procedure for obtaining the optimum slip load of FDDs in steel structures. In this research, by assuming identical optimum slip load for the FDDs in all stories, the normalized FDD strength was increased in a number of equal steps ranging from zero to one to determine the EFI of FDDs in each step. Then, the optimum slip load of FDDs in the steel frames was calculated according to the minimum EFI of FDDs. The findings of the present study are as follows:

- Since various ground motions provide various results to acquire the EFI curve and, accordingly, the optimum slip load of FDDs, to attain the best performance of FDDs in future earthquakes, the mean of results for ET and NTH analyses should be considered.
- The possibility of using the ET method to acquire the EFI and optimum slip load for FDDs was examined by comparing the results of ET and NTH analyses.
- The EFIs obtained from the mean of results for ET and NTH analyses were close to each other.
- As a result, utilizing the ET analysis instead of a large number of NTH analyses is also possible, and employing the ET method decreases 57% of computational efforts.
- A relation for calculating the optimum slip load of the FDDs in steel structures was proposed in terms of the weight of the structure.
- After adding optimal FDDs and investigating the effectiveness of dampers, it was concluded that by using ETEFs with better energy consistency, the ET results could be improved.